\documentclass[prl, twocolumn, superscriptaddress]{revtex4-1}
\usepackage{bm, amsmath, amsfonts, amssymb, braket}
\usepackage{subfigure}
\usepackage{color}
\usepackage{graphicx}
\usepackage{dcolumn} 
\usepackage{comment}

\usepackage{rotating} 

\begin{document}

\title{Non-Hermitian Skin Effects in Hermitian Correlated/Disordered Systems: Boundary-Sensitive/Insensitive Quantities and Pseudo Quantum Number}

\author{Nobuyuki Okuma}
	\email{okuma@hosi.phys.s.u-tokyo.ac.jp}
	\affiliation{Yukawa Institute for Theoretical Physics, Kyoto University, Kyoto 606-8502, Japan}
\author{Masatoshi Sato}
	\affiliation{Yukawa Institute for Theoretical Physics, Kyoto University, Kyoto 606-8502, Japan}

\date{\today}

\begin{abstract} 
There is a common belief in the condensed matter community that bulk quantities become insensitive to the boundary condition in the infinite-volume limit. Here we reconsider this statement in terms of recent arguments of non-Hermitian skin effects, -strong dependence of spectra on boundary conditions for the non-Hermitian Hamiltonians-, in the traditional Green's function formalism. We find the criterion for quantities to be sensitive/insensitive against the boundary condition in Hermitian correlated/disordered systems, which is characterized by the residue theorem.
We also discuss the uncertainty of the quasiparticle energy under the skin effects in terms of nonnormal pseudospectra, which can be tested via the sharp optical absorption from the bulk-surface coupling. Our result indicates that ``pseudo quantum number" emerges as a consequence of large nonnormality.

\end{abstract}

\maketitle
In the condensed matter community, bulk physics in solids has been believed to be insensitive to the boundary condition for a long time, which becomes a fundamental knowledge written in elementary textbooks \cite{Ashcroft-Mermin,Altland-Simons}. This assumption enables one to choose the periodic boundary condition (PBC) and use the Bloch wavefunctions for efficient bulk calculations, from thermodynamic/transport quantities \cite{Altland-Simons} to topological invariants \cite{Kane-review, Zhang-review}. 

Recently, eigenspectral properties of non-Hermitian Hamiltonians have turned out to be very different from the conventional Hermitian ones \cite{Bender-98, Bender-02, Bender-review, Konotop-review, Christodoulides-review, Gong-18, KSUS-19,KBS-19,Poli-15,Zeuner-15,Zhen-15,Zhou-18,Weimann-17,Xiao-17,St-Jean-17,Bahari-17,Harari-18,Bandres-18,Zhao-19,Brandenbourger-19-skin-exp,Ghatak-19-skin-exp,Xiao-19-skin-exp,Weidemann-20-skin-exp}. In particular, the non-Hermitian skin effect \cite{Hatano-Nelson-96,Hatano-Nelson-97,Lee-16,MartinezAlvarez-18,Torres-2019,YW-18-SSH,YSW-18-Chern,Kunst-18,YM-19,KOS-20,Lee-19,OKSS-20, Zhang-19,Okuma-19,Borgnia-19,Brandenbourger-19-skin-exp,Ghatak-19-skin-exp,Helbig-19-skin-exp, Hofmann-19-skin-exp,Xiao-19-skin-exp,Weidemann-20-skin-exp}, -strong dependence of spectra on boundary conditions for non-Hermitian Hamiltonians-, has been extensively studied. The spectral theory of the non-Hermitian skin effect has been sophisticated in terms of the boundary-localized modes called non-Bloch wavefunctions \cite{YW-18-SSH,YSW-18-Chern,Kunst-18,YM-19,KOS-20}, while the topological theories about its mathematical origin \cite{Lee-19,OKSS-20,Zhang-19} and symmetry-protected variants \cite{Okuma-19,OKSS-20} have also been developed.
Since the self-energy and the effective Hamiltonian modulated by the self-energy can be non-Hermitian in the Hermitian correlated/disordered systems,
these theories do not forbid the possibility of the skin effect even in the conventional Hermitian systems, which seems to contradict the assumption of the insensitivity against the boundary condition.
 
In this Letter, we reconsider the common belief of the boundary insensitivity by applying the theories of the non-Hermitian skin effects to the traditional Green's function formalism \footnote{While our study focuses on the skin effect of Green's function itself, Reference \cite{Borgnia-19} uses the Green's function as a mathematical tool to characterize the topological properties of the skin effect.}. We find the boundary-sensitive/insensitive quantities and characterize them in terms of the complex integral. We also find that the quasiparticle energy becomes ambiguous under the skin effect because it depends on a quantum measurement. We characterize this uncertainty by using the mathematics of the nonnormal pseudospectra and propose the notion of the pseudo quantum number.


\paragraph{Non-Hermitian Skin Effects of Green's Function.---}
Let us consider a one-dimensional bosonic/fermionic lattice system with $L$ sites and $m$ internal degrees of freedom. We assume the short-range interactions and the translation invariance, at least on average. 
Realistic two/three-dimensional materials can also be treated by regarding the momenta in additional dimensions as parameters.
If the total Hamiltonian consists of the quadratic terms of field operators and additional perturbation terms,
the retarded/advanced Green's function is defined as
\begin{align}
    \hat{G}^{\rm{R}/\rm{A}}(\omega)=\frac{1}{\omega-(\hat{H}_0+\hat{\Sigma}^{\rm{R}/\rm{A}}(\omega))},
\end{align}
where $\hat{H}_0$ and $\hat{\Sigma}^{\rm{R}/\rm{A}}(\omega)=[\hat{\Sigma}^{\rm{A}/\rm{R}}(\omega)]^{\dagger}$ are $Lm\times Lm$ matrices representing the unperturbed Hermitian Hamiltonian and the retarded/advanced self-energy induced by perturbations such as correlations and disorders, respectively.
For convenience, we define the effective Hamiltonian as $\hat{H}_{\rm eff}(\omega):=\hat{H}_0+\hat{\Sigma}^{\rm R}(\omega)$, which can be a non-Hermitian matrix. Regarding $\omega$ as a given parameter, one can apply various augments in non-Hermitian physics to this effective Hamiltonian \cite{Kozii-17,Yoshida-19,Michishita-20}.
Here we import the notion of the non-Hermitian skin effects, whose mathematical origin is the same as the exact zero modes of topological insulators/superconductors \cite{OKSS-20,Okuma-Sato-20}. 
In the case of one-dimensional systems with no symmetry (class A in Altland-Zirnbauer classification \cite{Altland}), 
if the spectral curve under the PBC has a nonzero winding in the complex energy plane, the class-A non-Hermitian skin effect inevitably occurs \cite{OKSS-20,Okuma-Sato-20}.
The simplest example is the Hatano-Nelson model without disorder \cite{Hatano-Nelson-96,Hatano-Nelson-97}:
\begin{align}
    [\hat{H}_{\rm HN}]_{i,j}=t(\delta_{i,j+1}+\delta_{i+1,j})+g(\delta_{i,j+1}-\delta_{i+1,j})-i 2g'\delta_{i,j},
\end{align}
with $t>0$ and $g'>|g|>0$, where $i,j$ denote the site indices, and the second term is a frequency-independent self-energy that represents non-Hermitian asymmetric hopping.
Although the realization of the second term is a nontrivial task in realistic materials, the same physics discussed in this Letter can be extended to skin effects induced by onsite self-energies \cite{YSW-18-Chern,KSU-18,Yi-Yang-20,Bessho-Sato-20}.
The third term ensures the negative imaginary part of the complex energy.
For large $L$, the PBC and OBC complex eigenspectra of $\hat{H}_{\rm HN}$ are given by
\begin{align}
    E_{\beta}=(t+g)\beta+(t-g)\beta^{-1}-i2g',\ 
    \beta=
    \begin{cases}
    e^{ik}& (\rm{PBC})\\
    re^{ik}& (\rm{OBC})
    \end{cases},
\end{align}
with $r=\sqrt{|t-g|/|t+g|}$, and $k\in[0,2\pi)$ is the crystal momentum.
In general, the OBC spectrum of a translation-invariant non-Hermitian Hamiltonian is calculated by the analytic continuation from the dispersion relation $E(e^{ik})$ to $E(\beta\in\mathbb{C})$, where the trajectory of $\beta$ called the generalized Brillouin zone $C_{\rm GBZ}$ is determined via the non-Bloch band theory \cite{YW-18-SSH,YSW-18-Chern,Kunst-18,YM-19,KOS-20}.
Since the right eigenvectors of $\hat{H}_{\rm HN}$ [i.e., $\hat{H}_{\rm HN}|\rm{r}\rangle=E|\rm{r}\rangle$] are given by a superposition of non-Bloch waves $|{\rm r},E_\beta\rangle_i=(\beta^{-i}-\beta^{*-i})/\sqrt{2L}$, the OBC eigenvectors are localized at one boundary for $r\neq1$, which is the origin of the name ``skin effect". In the following, we reveal what quantities are affected and unaffected by the skin effect.

\begin{figure}[t]
\begin{center}
　　　\includegraphics[width=8cm,angle=0,clip]{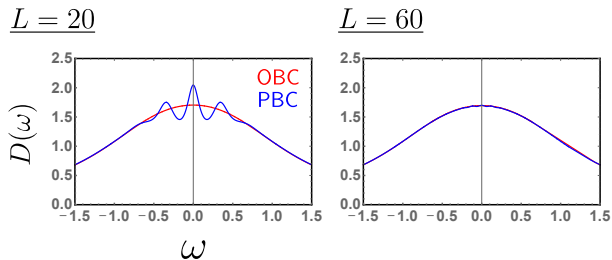}
　　　\caption{Density of states $D(\omega)$ of Hatano-Nelson model with $t+g=1$, $t-g=0.1$, and $g'=0.5$ under open (OBC) and periodic (PBC) boundary conditions. For large $L$, $D(\omega)$ becomes insensitive to the boundary condition.}
　　　\label{fig1}
\end{center}
\end{figure}

One of the boundary-sensitive quantities is the spectral function.
In the case of the Hatano-Nelson model, the spectral decomposition of the Green's function and the spectral function are given by 
\begin{align}
     \hat{G}^{\rm R}(\omega)=&\sum^{\pi}_{k=0}\frac{|{\rm r},E_\beta\rangle\langle {\rm l},E_\beta|}{\omega-E_\beta}=:\sum^{\pi}_{k=0} \hat{G}^{\rm R}(\beta,\omega),\\
     A(\beta,\omega):=&-2{\rm Im}\ {\rm Tr}\left[\hat{G}^{\rm R}(\beta,\omega)\right]
     =-2{\rm Im}\ \frac{1}{\omega-E_\beta},
\end{align}
where $\langle {\rm l},E_\beta|_i=(\beta^{i}-\beta^{*i})/\sqrt{2L}$ are the left eigenvectors of $\hat{H}_{\rm HN}$ [i.e., $\langle{\rm l}|\hat{H}_{\rm HN}=\langle {\rm l}|\hat{H}_{\rm HN}$]. Apparently, $A(k,\omega)$ is different from the spectral function under the PBC for $r\neq1$. 
For a given $k$, the peak and its half-width of the spectral function in frequency space characterize the energy and life-time of the quasiparticle. This implies that in general mode-resolved quantities are expected to drastically depend on the boundary condition.

Although the boundary condition affects the spectral property in the presence of the skin effect, 
it does not mean that the particles are localized at one boundary.
In fact, the particle density at site $x$ in equilibrium
\begin{align}
   n_x&={\rm Tr}\left[|x\rangle\langle x|\int_{-\infty}^{\infty}\frac{d\omega}{2\pi }f(\omega)i[\hat{G}^{\rm R}(\omega)-\hat{G}^{\rm A}(\omega)] \right]\notag\\
   &=\int_{-\infty}^{\infty}\frac{d\omega}{2\pi }f(\omega)\sum_{k}i\frac{\langle{\rm l},E_\beta|x\rangle\langle x|{\rm r},E_\beta\rangle    }{\omega-E_\beta}+c.c.,
\end{align}
where $f(\omega)$ is the equilibrium distribution function, is delocalized because the contributions from the right and left eigenstates localized at the opposite sides of the system cancel each other.

In addition to the particle density, mode-averaged quantities are insensitive against the boundary condition in the infinite-volume limit. First, we consider the density of states $D(\omega):=L^{-1}\sum_{k} A(\beta,\omega)$.
Figure \ref{fig1} shows the size-dependence of $D(\omega)$ of the Hatano-Nelson model calculated from a finite-size numerical diagonalization. For large $L$, $D(\omega)$ becomes insensitive to the boundary condition, which implies that the density of states does not depend on the boundary condition in the infinite-volume limit.
This statement can be easily shown for this simple model.
In the infinite-volume limit, the summation over $k={\rm arg}\ \beta$ can be replaced with the complex integral:
\begin{align}
    D(\omega)&=-2{\rm Im}\int^{\pi}_{0} \frac{dk}{\pi}\frac{1}{\omega-E_\beta}=-2{\rm Im}\int^{2\pi}_{0} \frac{dk}{2\pi}\frac{1}{\omega-E_\beta}\notag\\
    &=-2{\rm Im}\oint_{C_{\rm GBZ}} \frac{d\beta}{2\pi i}\frac{1}{\beta(\omega-E_\beta)},
\end{align}
where $C_{\rm GBZ}$ is the generalized Brillouin zone $r\mathbb{T}$ with $\mathbb{T}$ being the unit circle.
We use the fact that $k$'s are arranged at equal intervals, which is nontrivial in general non-Hermitian systems.
The final expression is applicable to the PBC case by replacing $C_{\rm GBZ}$ with the Brillouin zone $C_{\rm BZ}=\mathbb{T}$, which enables one to compare the OBC and PBC density of states in terms of the complex integral with residues that satisfy $\beta(\omega-E_\beta)=0$.  
Since all residues are placed outside the region between $\mathbb{T}$ and $r\mathbb{T}$, the integral does not depend on the boundary condition.

The above discussion can be extended to the general class-A non-Hermitian skin effect, and the following theorem holds.
\\

{\bf Theorem}~~Let $\sigma_{\rm PBC}[\hat{H}_{\rm eff}(\omega)]$ be the PBC spectrum of a class-A effective Hamiltonian under the infinite-volume limit ($L\rightarrow \infty$). If Im $\lambda<0$ for all $\lambda\in\sigma_{\rm PBC}[\hat{H}_{\rm eff}(\omega)]$, the density of states $D(\omega)=-2L^{-1}{\rm Im}\ {\rm Tr}[\hat{G}^{\rm R}(\omega)]$ does not depend on the boundary condition. 
\\

\begin{figure}[]
\begin{center}
　　　\includegraphics[width=8cm,angle=0,clip]{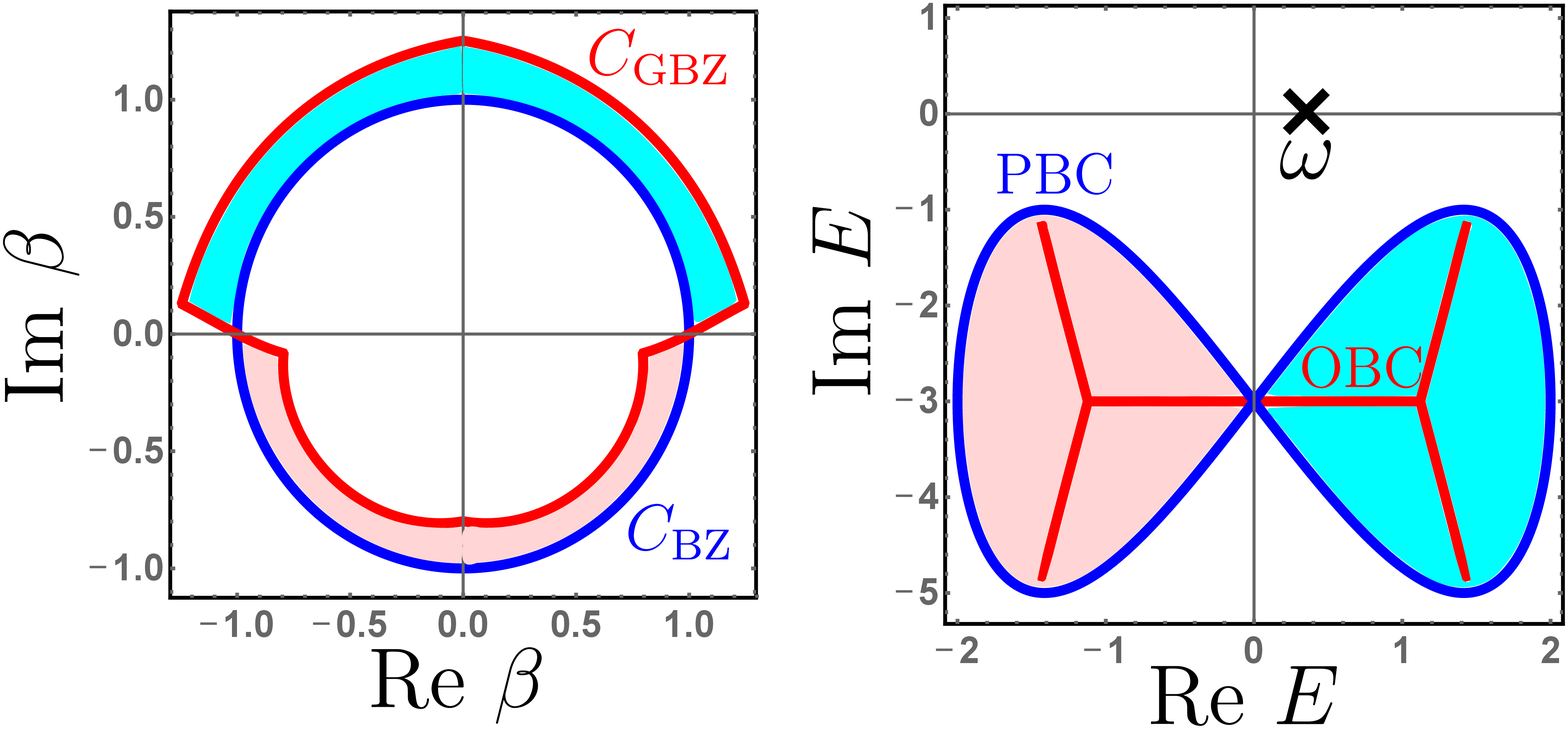}
　　　\caption{Relation between (generalized) Brillouin zone $C_{\rm BZ}$ ($C_{\rm GBZ}$) and PBC (OBC) spectrum. 
　　　The non-Bloch energy dispersion of this example (butterfly) is given by $E_\beta=\beta^2-i\beta+i\beta^{-1}-\beta^{-2}-3i$ \cite{Trefethen}.
　　　The shaded region of the $\beta$ space ($D_{\rm BZ}\triangle D_{\rm GBZ}$) corresponds to the shaded region of the complex energy space, which does not contain $\omega\in\mathbb{R}$.}
　　　\label{fig2}
\end{center}
\end{figure}

One can prove the Theorem by generalizing the above proof for the Hatano-Nelson model or using the knowledge of pseudospectra (see the latter part).
As for the former proof, there arise two subtle points. 
In general, the phase $\arg\ \beta$ is not guaranteed to be arranged at equal intervals under the infinite-volume limit.
This point is avoided by using the non-Bloch theory \cite{supplement}.

The other nontrivial procedure is to show that no zeros of $\beta(\omega-E_\beta)$ are in $D_{\rm BZ}\cap D_{\rm GBZ}$, where $D_{\rm BZ/GBZ}$ is the region surrounded by $C_{\rm BZ/GBZ}$.
For the proof, it is convenient to use the graphical discussion rather than the explicit expressions of zeros (Fig. \ref{fig2}).
According to the spectral theory \cite{Trefethen,Bottcher}, the symmetric difference $D_{\rm BZ}\triangle D_{\rm GBZ}$ in the $\beta$ space (shaded region in Fig. \ref{fig2}) corresponds to the region surrounded by the PBC curve with a nonzero winding number in the complex energy space, which does not contain $\omega$ on the real axis because of the assumption Im $\lambda<0$.
Thus, the zeros of $\beta(\omega-E_\beta)$ are not in that region, which indicates that the complex integral over $C_{\rm GBZ}$ is the same as that over $C_{\rm BZ}$.

The discussion based on the contour integral can be extended to other physical quantities.
In general, the skin effect of the effective Hamiltonian is not significant for a mode-averaged quantity if the integrand has no residue in $D_{\rm BZ}\triangle D_{\rm GBZ}$ because the path of the contour integral can be deformed from the generalized Brillouin zone to the conventional Brillouin zone without changing the result. In such a case, the only difference between the PBC and OBC quantities is the surface contributions that exist regardless of whether the skin effect occurs or not. Thus, lots of the equilibrium and non-equilibrium quantities that are expressed in terms of the mode-integration of the Green's functions are not affected by the skin effect. For example, we numerically check that the DC longitudinal conductivity is insensitive to the boundary condition \cite{supplement}.
Nevertheless, one can still expect detectable quantities affected by the skin effect. Next, we look more closely at the skin effect by introducing the notion of pseudospectra \cite{Trefethen,Colbrook-19,Okuma-Sato-20}. 

\paragraph{Uncertainty of Quasiparticle Energy and Pseudo Quantum Number.---}
In the presence of the skin effect of the Green's function, the definition of the quasiparticle complex energy becomes ambiguous in terms of the pseudospectrum, and such ambiguity can be observed as the uncertainty of an observable quantity in a realistic quantum measurement. We first apply spectral theory of a nonnormal matrix $\hat{H}$ (i.e., $[\hat{H},\hat{H}^\dagger]\neq0$) to the effective non-Hermitian Hamiltonian and then relate it to the measurement in the framework of the Green's function.

\begin{figure}[]
\begin{center}
　　　\includegraphics[width=8cm,angle=0,clip]{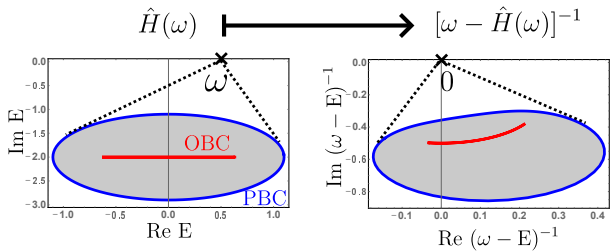}
　　　\caption{Pseudospectra of effective Hamiltonian and Green's function at $\omega=0.5$ of Hatano-Nelson model with $t+g=1$, $t-g=0.1$, and $g'=1$ under the limit $\lim_{\epsilon\rightarrow0}\lim_{L\rightarrow\infty}$. The shaded region corresponds to the boundary-localized non-Bloch waves. The physical condition Im$\lambda<0$ ensures the closed PBC spectral curve of the Green's function. }
　　　\label{fig3}
\end{center}
\end{figure}

In spectral theory of matrices, the $\epsilon$ pseudospectrum is a mathematical generalization of the spectrum under a small perturbation defined as the set below\cite{Trefethen,supplement,Okuma-Sato-20}
\begin{align}
    \sigma_{\epsilon}(\hat{H})=\{E \in\mathbb{C}|\ \|(\hat{H}-E)|v\rangle \|<\epsilon\ {\rm for\ some}\ |v\rangle\}.
\end{align}
In the case of normal matrices $\hat{N}$ (i.e., $[\hat{N},\hat{N}^\dagger]=0$) such as Hermitian matrices, the $\epsilon$ pseudospectrum is just the $\epsilon$ neighborhood of the spectrum, while it is larger than the $\epsilon$ neighborhood in the case of nonnormal matrices such as the Hatano-Nelson model under the OBC.
Physically, it means that small perturbations can induce a drastic difference under nonnormality.
In particular, the following relationship holds for a translation-invariant matrix $\hat{H}$ under the OBC \cite{Trefethen,Okuma-Sato-20}:
\begin{align}
    \lim_{\epsilon\rightarrow0}\lim_{L\rightarrow\infty}\sigma_\epsilon(\hat{H})=\sigma_{\rm SIBC}(\hat{H}),\label{sibc}
\end{align}
where $\sigma_{\rm SIBC}(\hat{H})$ is the spectrum under the semi-infinite boundary condition, in which there is only one open boundary. According to the index theorem \cite{Bottcher,Trefethen,OKSS-20,Okuma-Sato-20}, $\sigma_{\rm SIBC}(\hat{H})$ is given by the PBC spectrum together with the region enclosed by the PBC curve with nonzero winding, and the eigenvectors are the Bloch and non-Bloch waves with $\beta\in D_{\rm BZ}\triangle D_{\rm GBZ}$ (Fig.\ref{fig3}). The pseudoeigenvalues/vectors are given by these semi-infinite eigenvalues/vectors. 
Thus under the skin effect of the effective Hamiltonian in the infinite-volume limit, where the PBC curve has a nonzero winding, the pseudospectrum is given by the two-dimensional region including the OBC spectral curve even though $\epsilon$ is infinitesimally small.
This mathematical fact implies that one needs the perfect accuracy in the measurements to distinguish the spectrum from the pseudospectrum in solid-state physics, which is an unrealistic requirement.
Similarly, in the case of large but finite $L$, the pseudospectrum with $\epsilon$ being larger than an exponentially small threshold with respect to $L$ approaches asymptotically to $\sigma_{\rm SIBC}(\hat{H})$. Thus, for the given accuracy of the measurements, one cannot distinguish them in the system with sufficiently large size. 

The above mathematics can also be applied to the retarded Green's function by regarding it as a nonnormal matrix. Here we should assume Im$\lambda<0$ to ensure the closed PBC eigenspectral curve of the Green's function  (Fig.\ref{fig3}), which is needed for the index theorem.
In the following, we discuss the physics of the pseudospectrum under the skin effect in the level of a detectable quantity expressed in terms of the Green's function.
As mentioned above, the spectral function is a typical quantity that depends on the boundary condition.
In the absence of the skin effect, it is widely known to be a measurable quantity of the angle-resolved photoemission spectroscopy (ARPES) \cite{ARPES-review}. There an electron in the target band is excited by light into a reference (photo)electron state that has enough energy to escape from the solid, and the mode-resolved photoelectrons are detected as the photocurrent. Motivated by the ARPES measurements, we here consider the inter-band optical conductivity, which characterizes the ARPES photocurrent in the simplest approximation, between the correlated/disordered system under the skin effect and a reference state $|\alpha,{\rm ref}\rangle_i\propto \alpha^{-i}$ with $\langle \alpha,{\rm ref}|\alpha,{\rm ref} \rangle=1$.
In the linear response theory \cite{Fetter}, the real part of the inter-band optical conductivity, which describes the optical absorption rate, is given by 
\begin{align}
    {\rm Re}\ &\sigma^{\rm(inter)}(\Omega)\propto \frac{f(\xi_\alpha)-f(\xi_\alpha-\Omega)}{\Omega}\times\notag\\
    &\langle \alpha|\ i\left[\hat{G}^{\rm R}(\xi_\alpha-\Omega)-\hat{G}^{\rm A}(\xi_\alpha-\Omega)\right]\  |\alpha\rangle,\label{optcond}
\end{align}
where $\Omega$ is the photon frequency, and $\xi_{\alpha}\in \mathbb{R}$ is the energy of the reference state. $|\alpha\rangle_i\propto \alpha^{-i}$ with $\langle \alpha|\alpha \rangle=1$ is defined on the target system. In the case where the multi-band nature is important in the target system, the right-hand side should also contain the summation over the band indices, though it does not change the following discussion.
Here we assume that the Green's functions of the reference state can be replaced with the delta function because of the large lifetime. 

Although the exact eigenstates of the non-Hermitian effective Hamiltonian are characterized by $\alpha$ on the curve $C_{\rm GBZ}$, $|\alpha\rangle$ with $\alpha\in D_{\rm BZ}\triangle D_{\rm GBZ}$ is a pseudoeigenstate with the infinitesimally small $\epsilon$ in the infinite-volume limit. Correspondingly, it is also a pseudoeigenstate of the Green's function itself.
Thus, the matrix element of Eq. (\ref{optcond}) becomes
\begin{align}
    A_{\rm pse}(\alpha,\xi_\alpha-\Omega):=-2{\rm Im}\ \frac{1}{\xi_\alpha-\Omega-E_\alpha(\xi_\alpha-\Omega)},
\end{align}
where $E_\alpha(\omega)$ is the non-Bloch pseudoeigenvalue of $\hat{H}(\omega)$.
We define the pseudospectral function $A_{\rm pse}$ instead of the spectral function, which has the same form as the spectral function in the infinite-volume limit. Thus, the optical conductivity has the sharp peak even though it does not correspond to the exact complex energy of the effective Hamiltonian. If we set $\alpha=e^{ik}$ on $C_{\rm BZ}=\mathbb{T}$, the observed dispersion and lifetime are characterized by the PBC complex energy $E_k$. If we set $|\alpha|\neq1$, on the other hand, those are characterized by the non-Bloch energy $E_{\alpha}$, which means that the sharp optical absorption via the bulk-surface coupling is induced in the presence of the skin effect. Roughly speaking, the former/latter measurement with the plane-wave/non-Bloch-wave reference state mimics the ARPES measurement about real/complex momenta parallel/perpendicular to the surface \footnote{In the parallel direction, a photoelectron state is characterized by a Bloch wave.
In the perpendicular direction, on the other hand, a photoelectron state consists of a plane wave out of the solid and a damping wave with a complex momentum in the solid. For more sophisticated theories, see Ref. \cite{Hedin-02}}. 

The above arguments indicate that there is an ambiguity to determine the quasiparticle dispersion/lifetime owing to the nonnormal pseudospectral behavior. If one regards the photo-excitation process as a quantum measurement by a projection operator $|\alpha\rangle\langle\alpha|$, this ambiguity can be regarded as an emergent uncertainty of the measurement, which is absent in normal systems. While the quasiparticles in normal systems are labeled by a value $e^{ik}$ on the one-dimensional curve $\mathbb{T}$, those in the presence of the skin effect are labeled by a complex value $\alpha$ on the two-dimensional region $D_{\rm BZ}\triangle D_{\rm GBZ}$, which means that the large nonnormality in the infinite-volume limit effectively lifts the dimension of the good quantum number. The idea of the emergent quantum number would be an interesting direction of future works in non-Hermitian quantum physics with a small measurement error $\epsilon$ related to the nonnormal $\epsilon$ pseudospectrum.

The concept of the pseudospectrum can also be used to prove the Theorem, as noted above.
By inserting the unity $\hat{1}=\sum_{k}|k\rangle\langle k|$, where $|k\rangle:=|e^{ik}\rangle$ is a Bloch wave, to $D(\omega)=-2$Im Tr$[\hat{G}^{\rm R}(\omega)]$ under the OBC, we obtain
\begin{align}
    D(\omega)&=\frac{1}{L}\sum_{k}-2{\rm Im\ }\langle k|[\hat{G}^{\rm R}(\omega)]|k\rangle\notag\\
    &\xrightarrow{L\rightarrow\infty}-2{\rm Im}\int^{2\pi}_{0}\frac{dk}{2\pi}\frac{1}{\omega-E_k}.
\end{align}
In the second line, we use the fact that $|k\rangle$ is an pseudospectrum with infinitesimally small $\epsilon$. This expression is the same as $D(\omega)$ under the PBC.

We note that the relevance of the skin effect in the optical bulk-surface coupling is compatible with the fact that the integrand in the complex integral should contain the residues in $D_{\rm BZ}\triangle D_{\rm GBZ}$ for the relevance, which arises from the projection process onto $|\alpha\rangle\langle\alpha|$.
Mathematically, if we can omit the assumption ${\rm Im}\lambda<0$, lots of mode-averaged quantities such as $D(\omega)$ can also be affected by the skin effect because the zeros of $\beta(\omega-E_{\beta})$ can exist in $D_{\rm BZ}\triangle D_{\rm GBZ}$, which would be relevant for systems with amplifications such as classical systems.

Finally, we also note that our results about the class-A skin effect would be naturally extended to the other symmetry classes. For example, the skin effect of time-reversal-symmetric non-Hermitian Hamiltonians has been investigated in terms of the topological properties \cite{OKSS-20}, the non-Bloch theory \cite{KOS-20}, the infinitesimal instability \cite{Okuma-19}, and the pseudospectra \cite{Okuma-Sato-20}, which enable one to repeat a similar discussion in this Letter.

We thank Kohei Kawabata for a discussion about the simulation of non-Bloch bands. 
This work was supported by JST CREST Grant No.~JPMJCR19T2, Japan. N.O. was supported by KAKENHI Grant No.~JP18J01610 and JP20K14373 from the JSPS. M.S. was supported by KAKENHI Grant No.~JP20H00131 from the JSPS.

%

\widetext
\pagebreak

\renewcommand{\theequation}{S\arabic{equation}}
\renewcommand{\thefigure}{S\arabic{figure}}
\renewcommand{\thetable}{S\arabic{table}}
\setcounter{equation}{0}
\setcounter{figure}{0}
\setcounter{table}{0}

\begin{center}
{\bf \large Supplemental Material for 
``Non-Hermitian Skin Effects in Hermitian Correlated/Disordered Systems: Boundary-Sensitive/Insensitive Quantities and Pseudo Quantum Number"}
\end{center}

\section{SI.~Replacement of summation with integral}

We consider a general way to replace the summation over $\beta$ with an integral, which is more nontrivial than that for the Hatano-Nelson model.
Suppose that the OBC spectrum of $\hat{H}_{\rm eff}(\omega)$ is given by $E(\beta\in C_{\rm GBZ})$, where we omit the frequency and the band index for simplicity. According to the non-Bloch band theory \cite{YM-19}, $\beta$ satisfies the following-type relationship:
\begin{align}
    F_i(\beta(\theta))=F_j(\beta(\theta) e^{-2i\theta}),
\end{align}
where $F_i(x)$ is an algebraic function with finite degree whose detail is not important here, and $2\theta\in[0,4\pi)$ is the relative angle between two solutions that correspond to the same energy. In general, the quantity guaranteed to be arranged at equal intervals under the infinite-volume limit is not $\arg\ \beta$ itself but the relative angle $2\theta$ \cite{YM-19}.
Thus, the summation $1/L\sum_{E_\beta}(\omega-E_\beta)^{-1}$ is replaced with
\begin{align}
    \int^{\infty}_{-\infty}\frac{d\theta}{2\pi}\frac{1}{\omega-E_{\beta(\theta)}}=\oint_{\mathbb{T}} \frac{dz}{2\pi i}\frac{1}{z(\omega-E_{\beta(z:=e^{i\theta})})}
    =\oint_{C_{\rm GBZ}}\frac{d\beta}{2\pi i}\left[a(\beta)\frac{\beta}{z}\frac{dz}{d\beta}   \right]\frac{1}{\beta(\omega-E_{\beta})}.\label{integrand}
\end{align}
Here we determine the analytic continuation from $e^{i\theta}$ to $z\in\mathbb{C}$ via $F_i(\beta)=F_j(\beta z^{-2})$ and insert a non-singular function $a(\beta)$ that is unity on $C_{\rm GBZ}$. 
In the case of the PBC, $\arg\ \beta$ is arranged at equal intervals in contrast with the OBC, and the integrand is given by $[\beta(\omega-E_{\beta})]^{-1}$. By choosing $a(\beta)$ such that the square-bracket part of Eq. ($\ref{integrand}$) is unity on $\mathbb{T}$, one can express the integrand in the common form for both boundary conditions, as in the case of the Hatano-Nelson model. 

\begin{figure}[b]
\begin{center}
　　　\includegraphics[width=14cm,angle=0,clip]{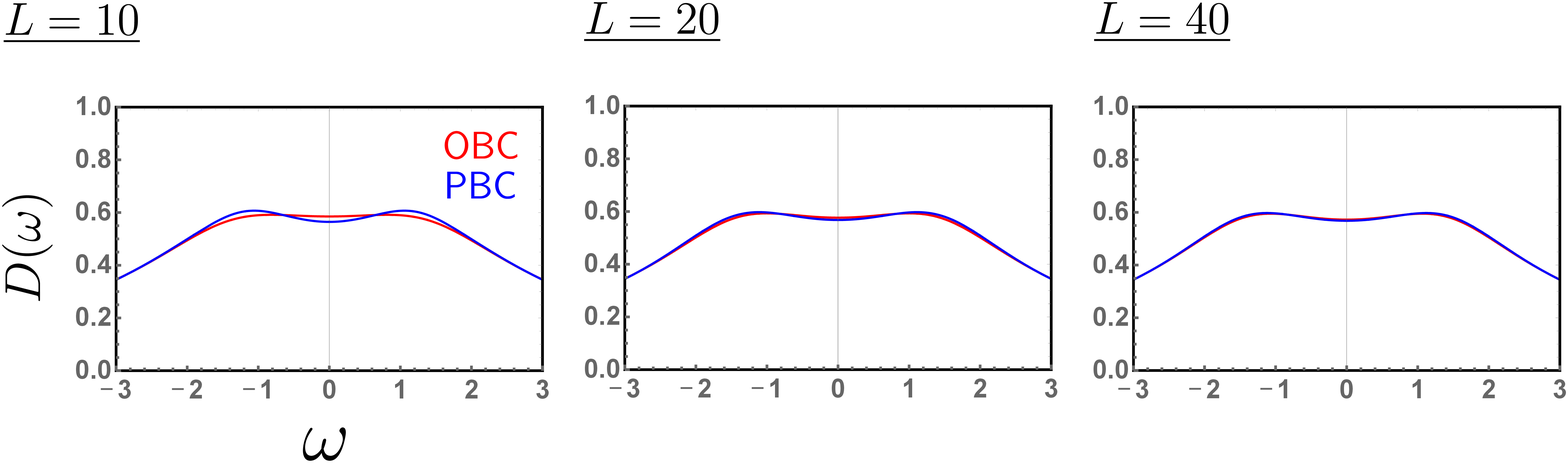}
　　　\caption{Density of states $D(\omega)$ of the butterfly model $E(\beta)=\beta^2-i\beta+i\beta^{-1}-\beta^{-2}-3i$ \cite{Trefethen} under open (OBC) and periodic (PBC) boundary conditions. For large $L$, $D(\omega)$ becomes insensitive to the boundary condition.}
　　　\label{s1}
\end{center}
\end{figure}

As an example of the non-Hermitian skin effect with a complicated generalized Brillouin zone, we plot the density of states $D(\omega)$ of the butterfly model in Fig. \ref{s1} (see also Fig. \ref{fig2}). As in the case of the Hatano-Nelson model, $D(\omega)$ is insensitive against the boundary condition for large $L$.

\section{SII.~DC longitudinal conductivity under skin effect}

\begin{figure}[]
\begin{center}
　　　\includegraphics[width=14cm,angle=0,clip]{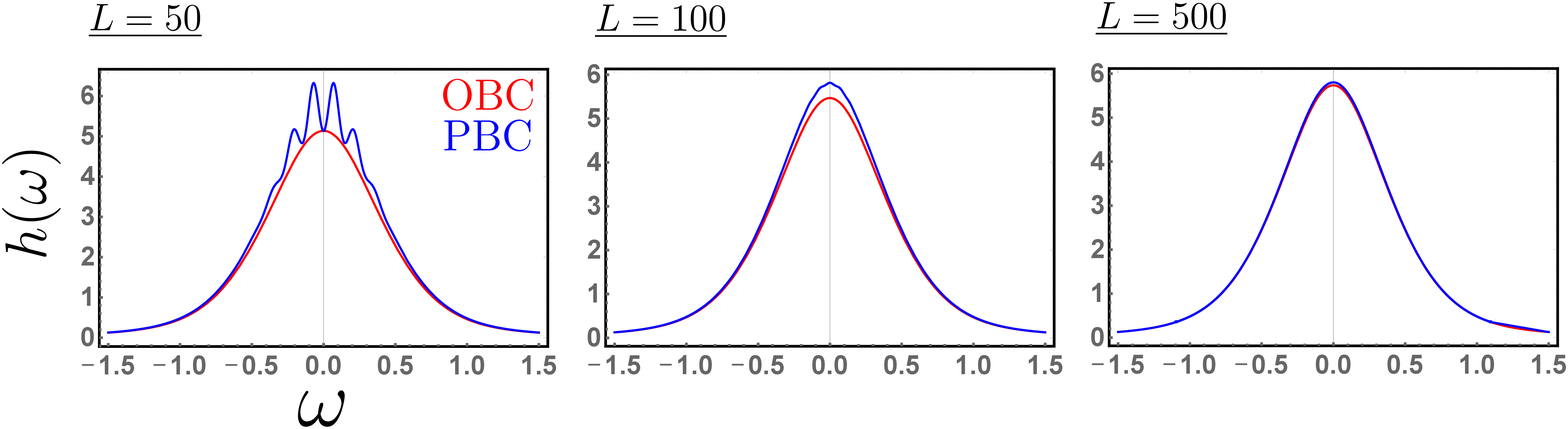}
　　　\caption{Size dependence of frequency-resolved contribution to DC longitudinal conductivity $h(\omega)$ of the Hatano-Nelson model with $t+g=1$, $t-g=0.1$, and $g'=0.5$. For large $L$, $h(\omega)$ becomes insensitive to the boundary condition.}
　　　\label{s2}
\end{center}
\end{figure}

We here evaluate the expectation value of the current operator $\hat{j}$ under a small electric field. 
We set the elementary charge as unity. 
In the linear response theory, the longitudinal conductivity without the vertex correction is given by
\begin{align}
\sigma_{xx}=-\int^{\infty}_{-\infty}\frac{d\omega}{2\pi}f'(\omega)\left[{\rm Tr}[\hat{G}^{\rm R}(\omega)\hat{j}\hat{G}^{\rm A}(\omega)\hat{j}]_{\omega}-\frac{1}{2}\left({\rm Tr}[\hat{G}^{\rm R}(\omega)\hat{j}\hat{G}^{\rm R}(\omega)\hat{j}]+{\rm Tr}[\hat{G}^{\rm A}(\omega)\hat{j}\hat{G}^{\rm A}(\omega)\hat{j}]\right)\right]=:-\int^{\infty}_{-\infty}\frac{d\omega}{2\pi}f'(\omega)h(\omega).
\end{align}
In the following, we perform the numerical calculation of the frequency-resolved contribution to DC longitudinal conductivity $h(\omega)$ of the Hatano-Nelson model. For the PBC calculation, we use the Fourier transformation and the momentum-resolved current $j_k=dH_0(k)/dk=-2t\sin k$, while for the OBC calculation, we use the real-space picture and the current operator $\hat{j}=1/i[\hat{r},\hat{H}_0]$ with the position operator $[\hat{r}]_{i,j}=i\delta_{i,j}$.
We plot the size dependence of $h(\omega)$ for both boundary conditions in Fig.\ref{s2}. For large L, $h(\omega)$ is insensitive to the boundary condition, which implies that the DC longitudinal conductivity does not depend on the boundary condition in the infinite-volume limit.

\section{SIII.~ Definition and properties of pseudospectra}
We here summarize the definition and the basic properties of pseudospectra.
For understanding of mathematical details, see Ref.\cite{Trefethen}.
For understanding of the relation between pseudospectra and topological insulators/superconductors, see Ref.\cite{Okuma-Sato-20}. 
\subsection{Definition of pseudospectra}
There are three identical definitions of $\epsilon$ pseudospectrum $\sigma_\epsilon(H)$ of a matrix $H\in \mathbb{C}^{N\times N}$ for arbitrary $\epsilon>0$:
\begin{itemize}
\item The set of $z\in\mathbb{C}$ such that $\|(z-H)^{-1}\|>\epsilon^{-1}$.
\item The set of $z\in\mathbb{C}$ such that $z\in\sigma(H+\eta)$ for some $\eta\in\mathbb{C}^{N\times N}$ with $\|\eta\|<\epsilon$.
\item The set of $z\in\mathbb{C}$ such that $\|(z-H)\bm{v}\|<\epsilon$ for some $\bm{v}\in\mathbb{C}^{N}$.
\end{itemize}
$\sigma(A)$ is the spectrum of a matrix $A$, and $\| A\|$ is the 2-norm of a matrix $A$ defined as
\begin{align}
    \|A\|:=\max_{\bm{x}}\frac{\|A\bm{x}\|}{\|\bm{x}\|}
\end{align}
where $\| \bm{x}\|$ is the conventional 2-norm of a vector $\bm{x}$.

\subsection{Properties of normal and nonnormal pseudospectra}
By definition, the pseudospectrum describes behaviors of spectra under perturbations.
In the case of Hermitian matrices, any perturbations to them do not change their spectra so much.
Actually, for general normal matrices (i.e., $[\hat{H},\hat{H}^\dagger]=0$), the pseudospectrum is given by the $\epsilon$ neighborhood of the spectrum \cite{Trefethen}:
\begin{align}
    \sigma_\epsilon(H)=\sigma(H)+\Delta_\epsilon:=\{z~|~{\rm dist}(z,\sigma(H))<\epsilon\},
\end{align}
where dist($\cdot,\cdot$) denotes the distance between two points in the complex plane.
In the case of nonnormal matrices (i.e., $[\hat{H},\hat{H}^\dagger]\neq0$), on the other hand, the pseudospectrum is larger than the $\epsilon$ neighborhood of the spectrum:
\begin{align}
     \sigma_\epsilon(H)\supset\sigma(H)+\Delta_\epsilon.\label{epsilonpseudo}
\end{align}
Equation (\ref{epsilonpseudo}) means that small perturbations to nonnormal matrices drastically change the spectrum.
In general, the upper bound of the pseudospectrum of a diagonalizable matrix is given in terms of the condition number $\kappa(P)$ \cite{Trefethen}:
\begin{align}
     \sigma(H)+\Delta_\epsilon\subseteq\sigma_\epsilon(H)\subseteq\sigma(H)+\Delta_{\kappa(P)\epsilon}.
\end{align}
The condition number is defined as 
\begin{align}
    \kappa(P):=\|P\|\|P^{-1}\|
\end{align}
where the matrix $P$ is defined as $H=PDP^{-1}$ with $D$ being the diagonal matrix whose elements are eigenvalues of $H$.
Since the 2-norm of a matrix is its largest singular value and the norm of the inverse is the inverse of the smallest singular value,
the condition number is calculated as
\begin{align}
    \kappa(P)=s_{\max}(P)/s_{\min}(P)\geq1.
\end{align}
This quantity measures the nonnormality of the matrix $H$ and becomes unity only when $H$ is normal.
When $H$ is not diagonalizable, $\kappa$ is set to be infinite as a convention.

\end{document}